# Cosmic magnetic fields and implications for HE particle anisotropies

Philipp P. Kronberg
*LANL, Los Alamos NM 87545, USA*
*and Dept. of Physics, University of Toronto, Toronto M5S 1A7, Canada*

I review what is known and surmised about magnetic fields in space, from our Milky Way environment to the distant Universe beyond the GZK horizon. This includes our gradually improving specification of the CR propagation environment within the Milky Way, the nearby universe within ~ 10 Mpc, and out to the GZK "horizon" near 100Mpc. Within these modest intergalactic distances we hope for some pointing capability for CR energies above ~ $10^{19}$eV, and for different species, as the observed event numbers accumulate in this range over the near future. The wider intergalactic propagation environment beyond the GZK horizon is also discussed. It sets a useful context for understanding other types of anisotropies, including sources of HE photons, neutrinos, leptons, etc. and for understanding relative time of arrival differences, such as those produced by lepton-photon cascades in the intergalactic medium. The global layout of potential UHECR sources is likely connected with the large scale structure (LSS) of cosmic filaments and voids, at least within ~ 100 Mpc. Possible source candidates for UHECR production are discussed, at various redshift ranges up to z ~ 2. Candidates discussed are AGN-jet sources, Centaurus A and more distant giant radio galaxies, and the possible indirect role of galaxies having a strong magnetized CR gas outflow that is driven by "starbursts" involving multiple supernovae and other energetic stellar events. Various analysis methods are described. I also discuss the current state of the results and near-future prospects for improving them.

## 1. INTRODUCTION

Deflection of CR's can happen on many places. Progressing outward, these are: (1) the Geo-magnetosphere, (2) the Solar System environment (3) the interstellar medium (ISM) in our vicinity of the Galactic (Milky Way) disk, (4) the Galactic halo and metagalactic environment, the (5) the intergalactic environment within a few Mpc, and out to Centaurus A at 3.8Mpc, (6) extragalactic space out to approximately a GZK radius, (7) the larger z Universe out to $z \approx 2$.

In this presentation I ignore zones (1) and (2), and begin with the Milky Way's disk's environment. Here the general magnetic field strength is typically 2 – 5$\mu$G and the local Galactic spiral arm cross section dimension is ~ 2 – 3kpc. This curved, "magnetized tube" extends over ~ 10kpc in the plane of the Galactic disk. However within the disk there are many magnetized gaseous systems such as old supernova remnant (SNR) shells and HII regions. Fields within these systems are locally much higher than 5$\mu$G.

Beginning at 100Mpc, that is in distance "zone" (7) we encounter large galaxy clusters. These are the largest gravitationally bound systems in the Universe, and magnetic fields their intra-cluster medium, (ICM), ~ 1Mpc in total size, are relatively well-measured at ~ 3 – 5 $\mu$G. This is comparable with the Milky Way disk, but in thermal ionized plasma densities of $10^{-2}$ – $10^{-3}$ cm$^{-3}$, which is 2 – 3 orders of magnitude less than in the Milky Way disk. Significantly, the ICM energy densities are comparable to those of the ISM, because the ICM gas temperature is 2 – 3 orders of magnitude higher than that of the ISM.

Galaxy clusters are less relevant for our study of UHECR's for two reasons. (a) the gyroradii of UHECR protons in the ICM is less than the cluster radii, so that they remain captured within the cluster while losing much of their energy through radiation and *p*-*p* collisions with thermal ICM protons. (b) Galaxy clusters occupy a very small fraction of the intergalactic volume, so they are seldom encountered by HE particles and photons propagating through the general intergalactic medium (IGM).

Except for the galaxy clusters, less is known about the magnetic field strength and structure in (4) – (7) above. However probes are continuously improving, and these are discussed here. Also, HE photons and CR's themselves can probe the intergalactic magnetic field strength, for example though particle-photon cascades, propagation time delays, and measurable particle deflections.

In the following I give an overview of what we know about magnetic field strengths in the various distance régimes (3) – (7) mentioned above.

## 2. MAGNETIC FIELD STRUCTURE ENVIRONMENT IN THE MILKY WAY

### 2.1. Global field structure from RM's

The likely global structure of the magnetic field in our Galaxy can be gleaned from two examples, shown below, Messier 51(NGC5194) seen face-on, and NGC891, an edge-on example. Both were imaged at more than one radio frequency, so that the local projected plane of polarization is corrected for Faraday rotation ($\propto \lambda^2$) to its intrinsic orientation. Figure 1 shows a large scale ordered magnetic field component that is approximately parallel to the spiral arms. The edge-on image of NGC891 in Figure 2 gives a projected view of the above-plane magnetic structure, which has a characteristic X-shape. The magnetic field lines are directed vertically outward above the nuclear zone, and gradually become more parallel to the disk at larger galactocentric radii.

To calculate the HECR propagation paths to us, especially below ~ $10^{18}$eV, we need to know the 3-D magnetic structure of the Milky Way, and any superimposed magnetic turbulence. The 2-D images in Figs 1 and 2 give important clues on the Milky Way's projected global field pattern. Galactic dynamo calculations have predicted either a dipole or quadrupole field structure, and these have been so far been difficult to verify observationally in any nearby galaxy, including the Milky Way. There is some evidence to show that neither a pure dipole or quadrupole give an ideal fit to the data, but one or the other seems to give a good approximation in some restricted investigations. The 3-D off-plane structure of the Milky halo has been difficult to discern observationally because there is relatively little Faraday rotation at high Galactic latitudes ($b$), hence at high $z$-heights above the plane. Models for our Galaxy's magnetic structure are best based on an all-sky distribution of the RM's of extragalactic radio sources, and/or of (mostly Galactic) pulsars. Pulsar RM's to date have been smaller in number than their e.g.r.s. counterparts, and their RM pathlengths through the ISM cover only part of the extragalactic source pathlength. An advantage of pulsar magnetic probes is that the pulsar dispersion measures (DM) can be used, via an ISM model of $n_e$, to model the *RM path length* through the Milky Way disk. Future larger numbers of RM and DM for the pulsars may lead to the most definitive 3-D Galactic disk field models for propagation paths and deflections of HE charged particles.

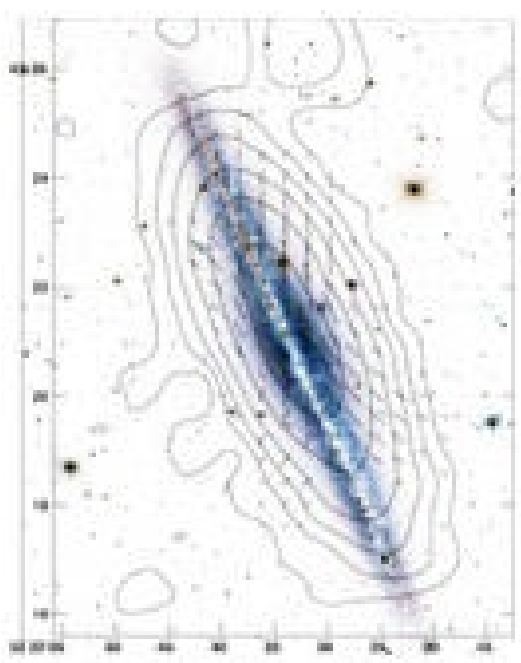

Fig. 1. NCC891 imaged with the Effelsberg radio telescope, with Faraday RM-corrected, projected magnetic field orientations [1] . *By permission of R. Beck.*

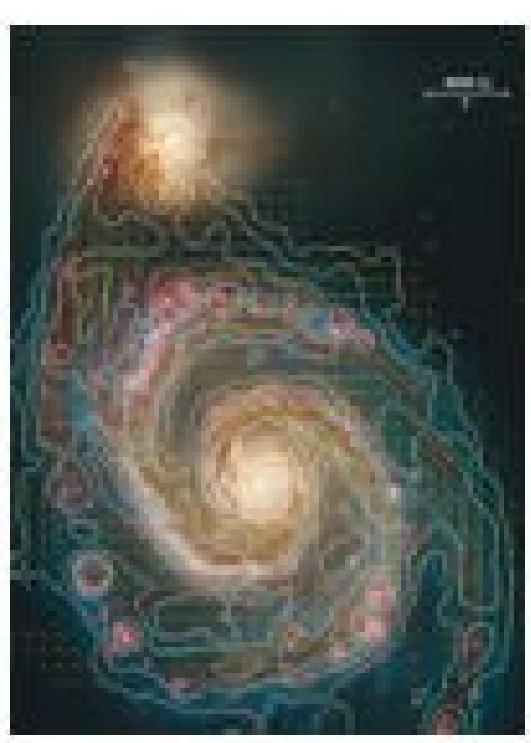

Fig. 2. Face-on galaxy M51, showing an overlay of the optical image, radio contours, and the RM-corrected magnetic field orientations as in Fig. 1. [2]. *Reproduced from Sterne und Weltraum by permission of R. Beck.*

Meanwhile, recent progress has been made with ever larger numbers of extragalactic source RM's, as is illustrated in a recent all-sky plot of *ca* 2250 rotation measures from a recent compilation made in 2009 [3]. A 3-panel plot in Figure 4 shows the clearest demonstration yet for an underlying magnetic field in our region of the Milky Way disk. This brings us to the most immediately relevant zone for calculating corrected UHECR arrival directions from outside the Galaxy.

The smoothed RM plot in Figure 4 is derived from 1500 newly determined RM's, plus *ca*. 750 additional RM's published in the literature. The 1500 set are an extension of the 555 RM set of Simard-Normandin, Kronberg & Button and were determined by the same 7-step procedure described in that paper [5]. They were then smoothed using a procedure developed by Simard-Normandin & Kronberg [4], which iteratively tests for, and rejects RM outliers, and/or unreliable RM values. These include genuinely anomalous RM's that deviate from the "RM consensus" in a given ($l$,$b$) direction.

Using this new compilation, Kronberg & Newton-McGee [3] produced plots, shown in Figure 5, of similarly smoothed RM's for different subsets of only low Galactic latitude RM's – in this plot only for RM's at latitudes $|b| \leq 15^{\circ}$. The smoothing full width was close to 20º, and this is near to ideal for "filtering out" an underlying coherent component of the Galactic disk field. These results, together with the higher latitude RM's in Figure 3 are being used with other input to construct propagation models to match with UHECR observations.

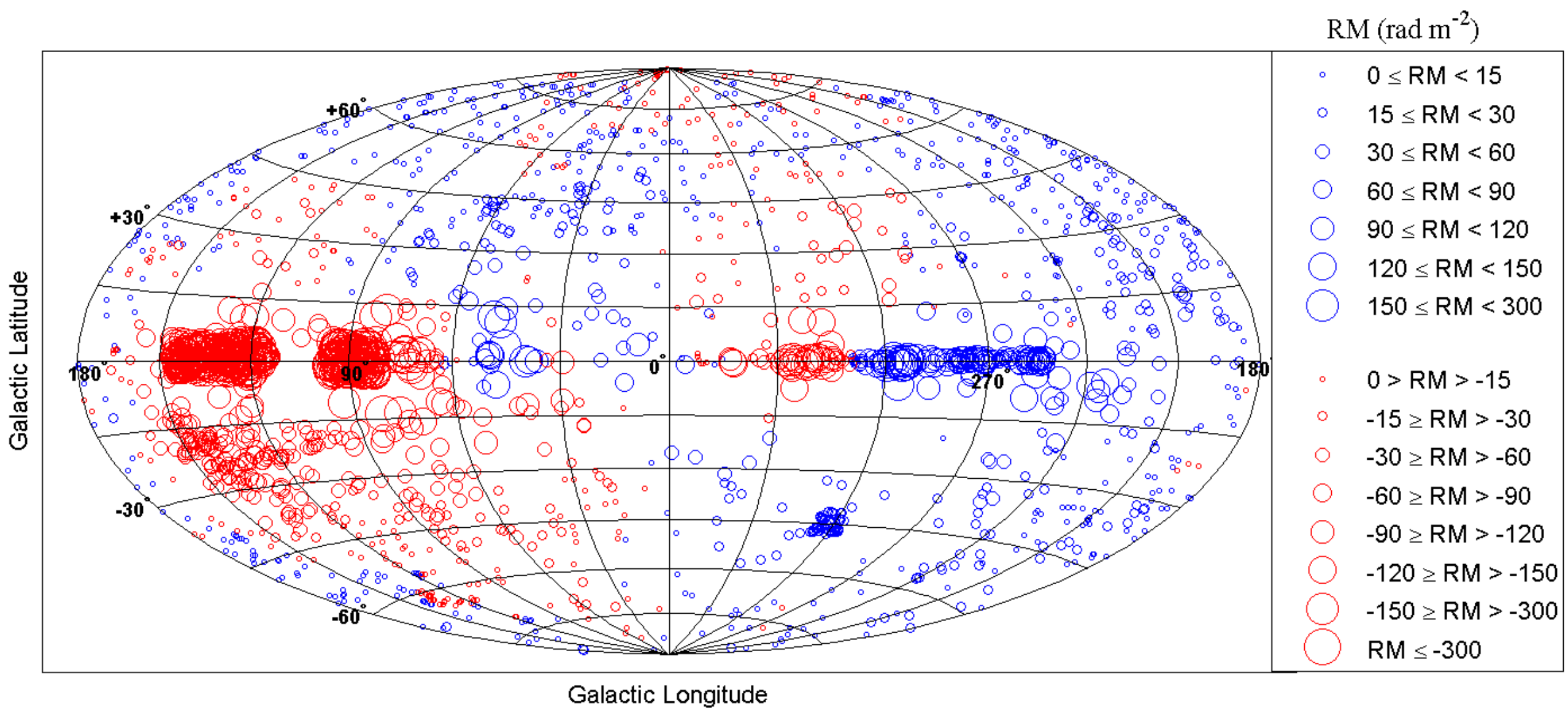

Figure 3: A smoothed representation of 2257 Faraday rotation measures in Galactic coordinates with the Galactic centre at (0,0). (Kronberg & Newton-McGee, [3]). Blue and red circles represent positive and negative RM's respectively, and the circle size is proportional to RM strength.

(a) (b) (c)

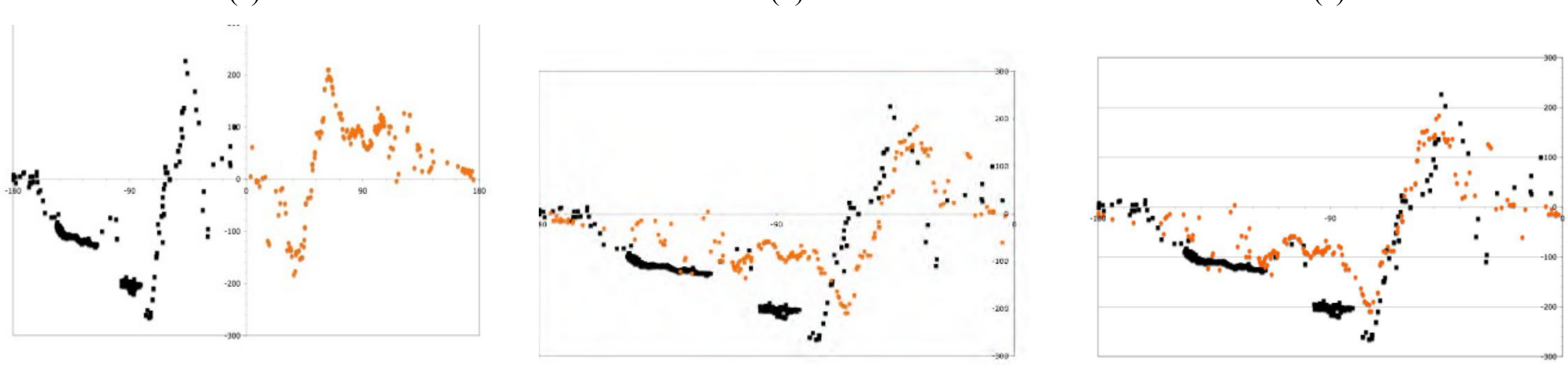

Figure 4. (a)Variation of RM's over the Galactic plane, when smoothed with a resolution roughly comparable with the width and (height) of a spiral arm. Black and orange identify RM's on either side of $l = 0°$. (b) The orange points are folded about $l = 0°$, and their RM sign is reversed (*i.e.* fold, reverse). (c) The difference between the curves in (a),(b) at $|l| \leq 90°$ for a range of shift angles was minimized to obtain a least squares minimized, optimum relative shift of $11° \pm 2°$ (fold, reverse, shift).

## 2.2. |$B_t$| vs. galactocentric radius

Another all-sky investigation has used the 408MHz all sky continuum synchrotron radiation survey of Haslam et al [6] and a more recent 1.4 GHz survey by Reich & Reich [7]. Breuermann et al [8] and more recently Berkhuijsen [9] have combined these data with various modeling assumptions to estimate the variation of total magnetic field strength, |$\boldsymbol{B}_t$|, with Galactocentric radius in the disk plane. This is shown in Figure 5. It shows an approximately exponential decay form

$$B_{\mathbf{t}}(R) = B_{\mathbf{0}} e^{\frac{R}{R_{\mathbf{0B}}}} , \qquad (1)$$

where $\boldsymbol{R}_{0B}$ = 11 ± 0.4 kpc. We note from Fig. 5 that |$\boldsymbol{B}_t$| remains a surprisingly strong 4 $\mu$G at $R$ = 17kpc. The results in Fig. 5 give good overall agreement with similar results from Broadbent *et al*. [10] and more recent study by Strong *et al* [11]. The latter are based on $\gamma$-ray emission and, notably, they are independent of the CR/magnetic energy equipartition assumptions needed in the radio sky analyses.

These results lead us into the less determined metagalaxy zone, or interface to the magnetic structure of the IGM in the local Universe (Zone (5))

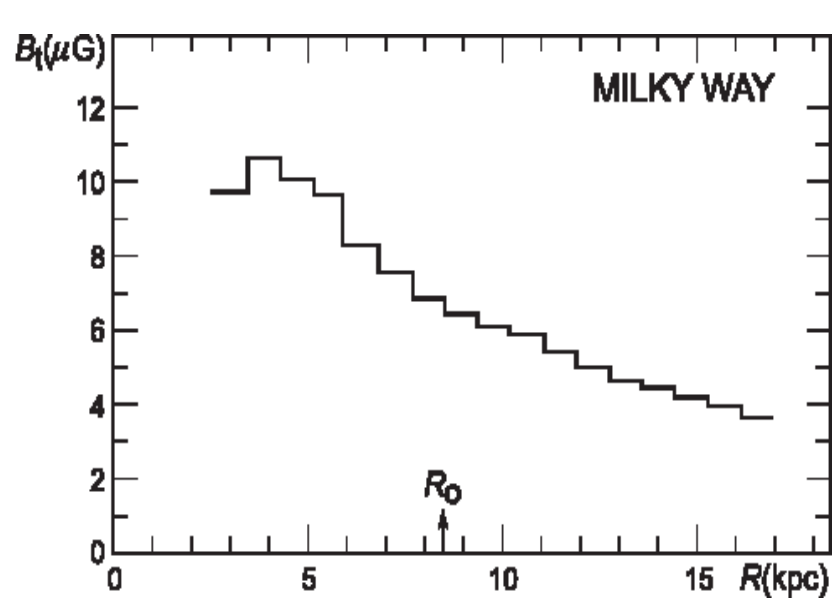

Figure 5 The variation of modeled interstellar magnetic field strength, |$\boldsymbol{B}_t$|, as a function of Galacto-centric radius out to 17 kpc from Berkhuijsen [9]. See text for further details. *Reproduced with permission from Elly Berkhuijsen.*

## 3.0 Centaurus A and candidate UHECR sources

### 3.1 The case of Centaurus A (NCG5128)

The interesting case of Centaurus A has been much discussed because of its proximity at 3.8Mpc, much less than a GZK radius, and the statistical likelihood of its coincidence with a slight excess of AUGER events within a few degrees – see Figure 6. Also it belongs to a category of AGN-powered radio sources that are plausible candidates as UHECR acceleration sites.

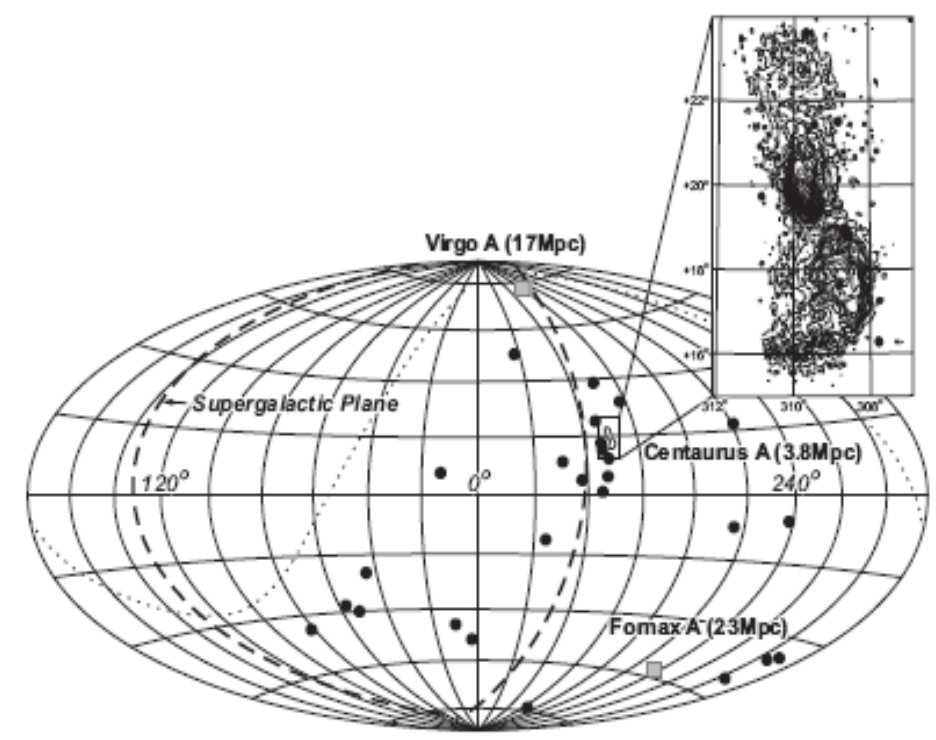

.

Figure 6 A plot in ($l$,$b$) of the CR events above $5\times10^{19}$eV as of 2007 [12]. Dashed lines show the supergalactic plane, and the inset shows a $\lambda$6.3cm image of the giant jet-lobe radio source Centaurus A [13], in a version kindly provided by Patricia Reich.

If this association is real, the scatter of arrival directions of a few degrees around the Cen A position is entirely plausible, given the expectation of the propagation path – which includes the environment of Cen A itself, ~ 3.5Mpc of intergalactic pathlength from Cen A, and finally the local Milky Way environment.

For small angle deflections $\theta(E)$ over a path $l$ containing $\boldsymbol{B}$ fluctuations on a scale of $l_0$(« $l$) and a UHECR nucleus of charge $Z$, the deflection is

$$\theta \approx 8^\circ Z \left(\frac{l}{10\ \mathrm{Mpc}}\right)^{0.5} \left(\frac{l_0}{1\ \mathrm{Mpc}}\right)^{0.5} \left(\frac{E}{10^{20}\ \mathrm{eV}}\right)^{-0.5} \left(\frac{B}{10^{-8}\ \mathrm{G}}\right)^{0.5} \qquad (2)$$

Inserting $l$ = 3.8 Mpc, $\boldsymbol{B}$ of $10^{-7}$ – $10^{-8}$ G, $E = 10^{20}$ $Z$ = 1, and a guess of $l_0$ ~ 0.3 Mpc brings us to a likely deflection range of a few degrees. Adding *Fe* into the mix of CR nuclei would greatly increase $\theta$ for the iron events, given the other assumed parameter values adopted above. Therefore, concentration of events within 10º - 20º of Cen A's direction would seem to rule out a 100% *Fe* contribution.

Recent Faraday RM data for the Milky Way suggest a very small RM in the Galactic halo above our disk location of 8kpc from the Galactic center (*e.g.* [3] and Mao *et al.* [14]). A recent RM study of the angular vicinity of Cen A (Feain *et al.* [15]), in which Cen A's outer radio contours will be recognized from Fig. 6), finds at most a very small RM contribution from the immediate intergalactic environment of Cen A (Figure 7). The fact of this small RM perturbation around the Cen A direction may be encouraging for the interpretation of UHECR arrival directions, given that there are no strong deflections beyond what we might expect from the Milky Way. The MW component probably comes from a propagation path mostly within ~ a disk $z$-height of ≈ 3 kpc [3]. All of this assumes that Cen A (and other AGN BH jet-lobe sources) are indeed sources of UHECR's. The next section briefly examines evidence for this possibility.

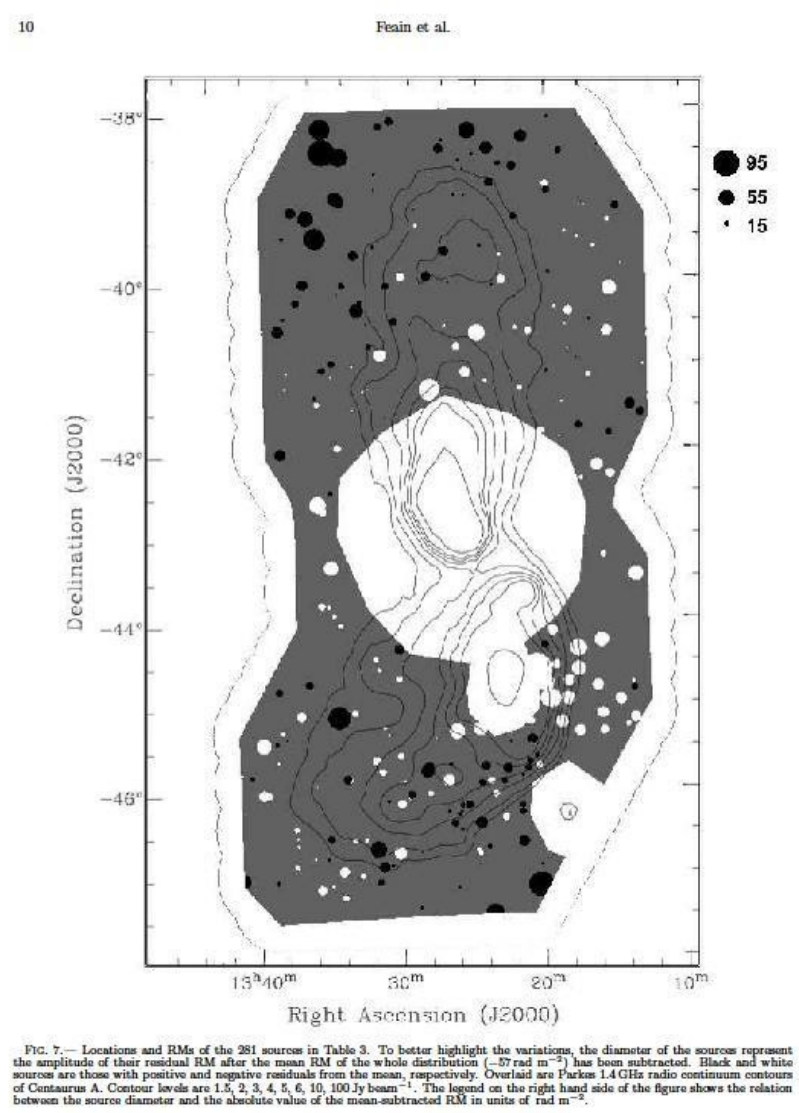

Figure 7 Background RM's in the direction of CenA in rad $m^{-2}$, from [15], including their original caption. Positive and negative RM values are shown black and white, respectively.

## 3.2: 3C303 – a possible UHECR acceleration site

The magneto-plasma parameters of the radio and X-ray jet emission knots of 3C303, a radio galaxy at $z$ = 0.141, have been diagnosed from multi-frequency observations [16]. Figure 8 shows three radio (and X-ray) emitting ``knots'' over a total distance of 28kpc. Each knot's projected magnetic field structure is remarkably coherent, with an average projected magnetic field direction close to the jet axis. The visible "cocoon" around the jet is ~ 1 kpc thick, and contains a Faraday-rotating, non-relativistic electron component whose plasma $\beta$ is ~$10^{-5}$ $T_8$ ($T_8$ in units of $10^8$ K), so that its Alfvén speed is in the relativistic régime. $|\boldsymbol{B}|$ in the cocoons is ~ $3\times10^{-3}$ G at ~ 0.5 kpc from the jet axis. Combining this with the 2 kpc knot length on a Hillas plot suggests that 3C303's jet knots are potential acceleration sites for CR nuclei up to ~ $10^{19\text{-}20}$ eV. That is,

$$E = \frac{B}{3\text{ mG}} \times \frac{L}{1\text{ kpc}} \quad \Rightarrow 2\times\ 10^{19}\ \text{ eV} \tag{3}$$

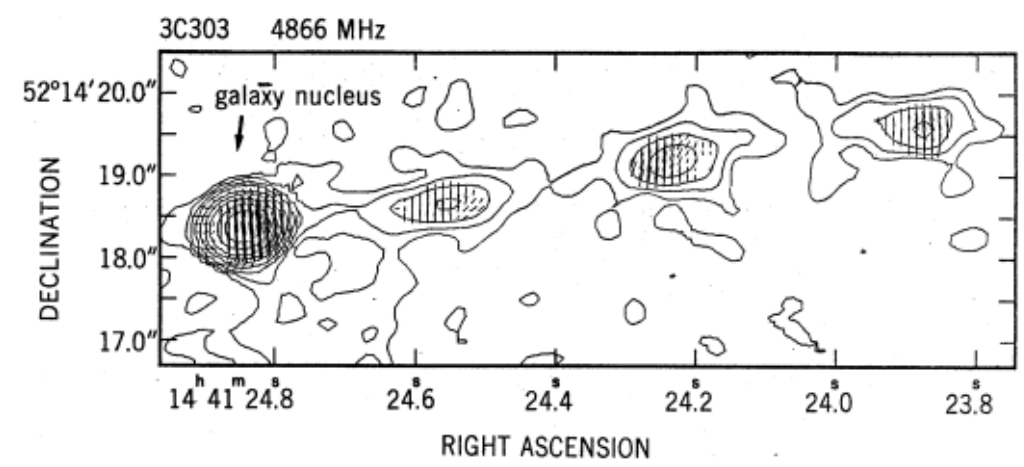

Figure 8. A 5 GHz VLA image of the main jet in 3C303 at 0.3" resolution. Lines show the Faraday rotation-corrected intrinsic polarization direction (i.e. normal to the projected B-orientation).

The energy flow along the jet is ~$3\times10^{43}$ $\tau_7^{-1}$ ergs $s^{-1}$ ($\tau_7$ is the total outflow time in units of $10^7$yr)[16]. The jet current has been determined at ~$10^{18}$ ampères along one of the knots [17]. This is the first direct observational estimate of extragalactic jet current, and is probably the largest current yet measured. Although the detailed mechanism for accelerating the CR's is not specified by the above, it would probably be in a $j_{\parallel}$ configuration in the highly aligned field by a current carrier starvation process [18], rather than a "conventional" Fermi-type shock mechanism. The expected natural fall-off of field strength $\boldsymbol{B}(r)$ --- $\propto r^{-1}$ — with transverse distance from the invisible, thin energy "pipe" in Figure 8 implies that the above reference value of ~ 1mG is location-variable within the jet cocoon.

## 4.0 Global evidence for magnetic energy in intergalactic space

### 4.1 Black holes as sources of IGM magnetic fields

The very largest BH-fed radio sources appear to have lost the least energy in other forms (e.g. PdV expansion work) as they expand into space, and in this sense are the best quantitative calibrators of gravitational-to-magnetic energy in the IGM[19]. Figure 8 illustrates this with a plot of the estimated energy content of giant radio sources. It is less than 2 orders of magnitude below the gravitational infall energy (to the Schwarzschild radius) of a $10^9$ $M_\odot$ black hole. Various corrections discussed in [19], such as particle diffusion etc. would correct the GRG points upward, and the BH energy downward, thus further narrowing the energy gap. The gap between the cluster sources and the giant radio lobes is approximately the independently measurable PdV work done as the lobes expand against the pressure of the ICM.

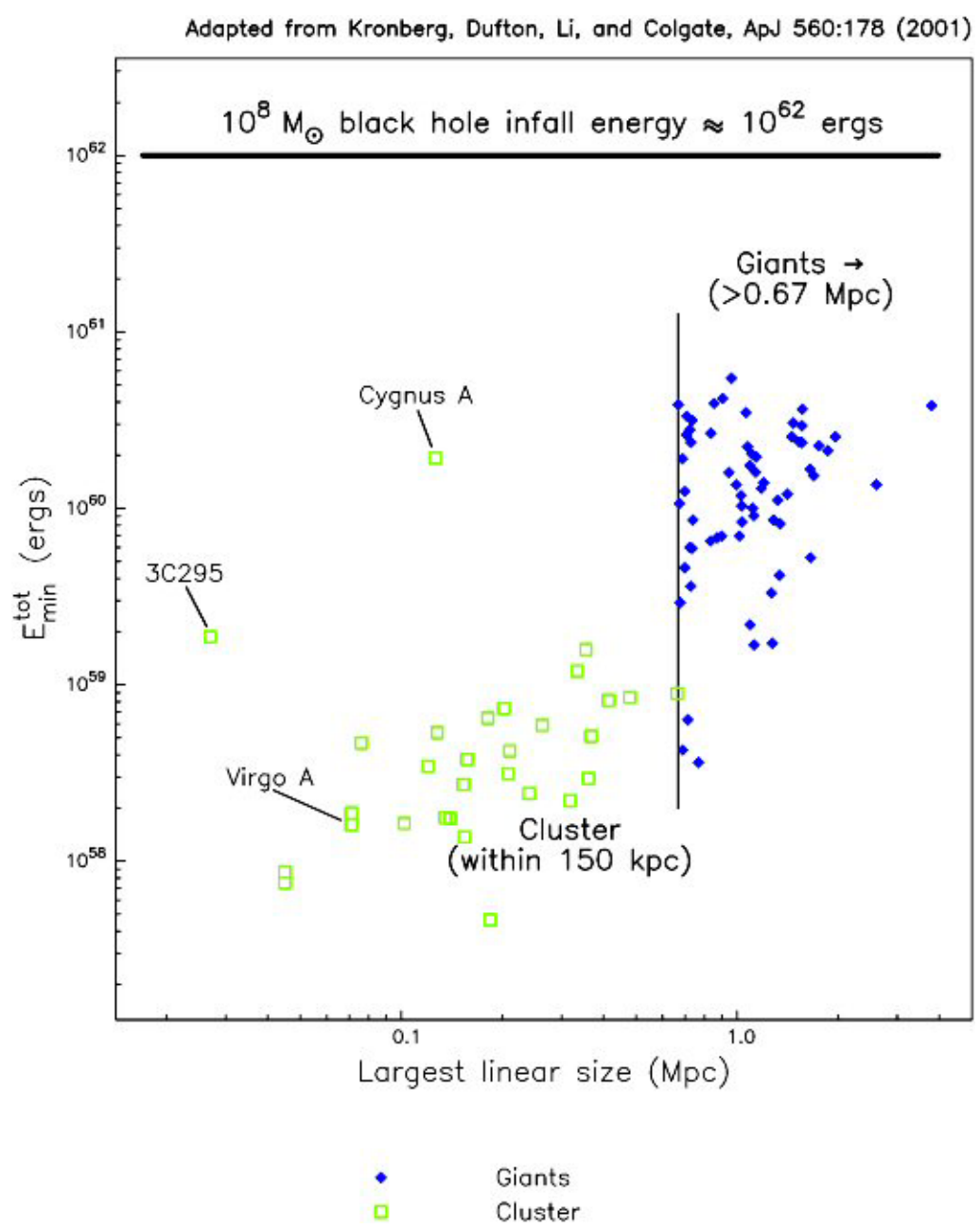

Figure 9 Radio lobe energy content for extended sources within galaxy clusters (squares) and the very largest BH-powered radio sources (diamonds) showing that the upper envelope of the latter is within 2 orders of magnitude of the putative gravitational formation energy of the central supermassive black hole [19].

The high efficiency of the gravitational to magnetic energy conversion implied by Fig 8, combined with the known space density of > $10^6$ $M_\odot$ black holes, ~ $10^5$ $M_\odot/\mathrm{Mpc}^3$, implies a global magnetic energy density in the galaxy over-dense zones (the galaxy filament zones of LSS) which is

$$\varepsilon_B = 1.36\times10^{-15}\left(\frac{\eta_B}{0.1}\right)\times\left(\frac{f_{RG}}{0.1}\right)\times\left(\frac{f^{VOL}{}_{\mathrm{FILAMENTS}}}{0.1}\right)^{-1}\times\left(\frac{M_{BH}}{10^8 M_\odot}\right)\ \mathrm{erg\ cm^{-3}} \quad (4)$$

$\varepsilon_B$ is the intergalactic energy density, $\eta_B$ is gravitational to magnetic energy conversion efficiency factor, $f_{RG}$ the fraction of all $L^*$ galaxies that produce radio lobes over a Hubble time, and $f^{VOL}{}_{\mathrm{FILAMENTS}}$ is the volume fraction of the mature universe (still well beyond a GZK distance) that is occupied by LSS filaments, *i.e.* the complement of the cosmic void fraction. For the normalizations adopted in (4), the corresponding intergalactic magnetic field strength is

$$B^{BH}_{IG} = \sqrt{8\pi\varepsilon_B} = 1.8\times10^{-7}\ \ \mathrm{G} \quad (5)$$

Before describing some first attempts below to detect and measure IGM fields, I now turn to another, though less quantifiable source of intergalactic fields due to star-driven outflows.

## 4.2 Early dwarf galaxy outflows before *z* ~7 as sources of IGM magnetic fields

Using detailed starburst outflow parameters measured for a mass range of galaxies down to ~ $10^8$ $M_\odot$ it is possible to project such data backwards in Cosmic time to $z \approx 10$, where the ~1000x smaller co-moving volumes might have contained mostly dwarf galaxies, since hierarchical merging into larger galaxies will have taken place mostly after that time. At $z \approx 10$, each co-expanding "cell" will be more easily filled with star and supernova-driven magnetized winds.

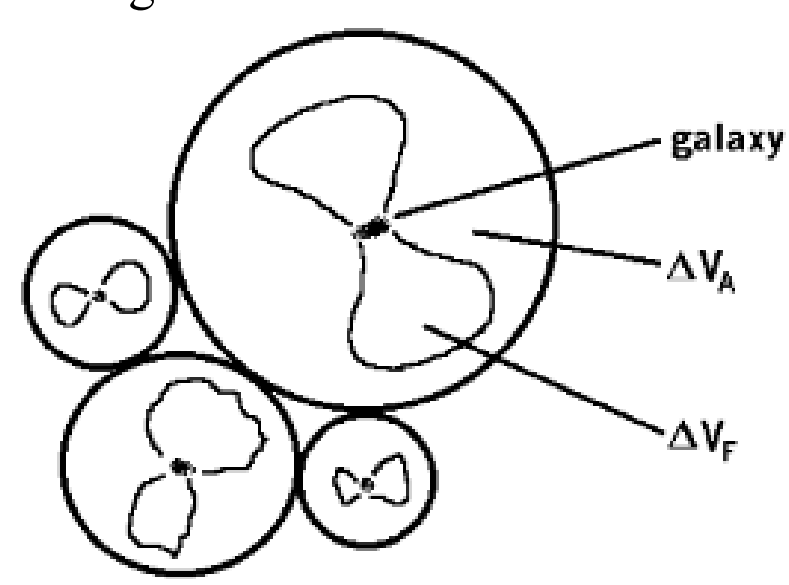

Figure 10 A cartoon dwarf galaxy outflow filling of the IGM. $\Delta V_F$ is the local IGM volume filled by the starburst outflow, and $\Delta V_A$ the volume available to be filled, in a bubble model representation of intergalactic space (ref. [20]).

As time progresses toward the present epoch, these co-expanding cells will retain their volume filling factor from early times when they were small in absolute size. Model calculations of this kind [20] show two interesting results that persist over a wide range of model parameters: First, the global fraction, $\Sigma\ [\Delta V_F]$ in Fig. 10, of IGM volume within the galaxy filament zones that is ultimately filled with dwarf galaxy outflow magneto-plasma reaches ≈ 20% of $\Sigma\ [\Delta V_A]$ by the present epoch ($z = 0$). Second, in a variety of model parameter combinations this substantial filling factor nearly reaches the $z = 0$ value by $z \sim 7$ [20].

If the "$\Delta V_F$ clouds" were to expand adiabatically as Proper Time proceeds, this collective starburst outflow contribution to $<|\boldsymbol{B}_{IGM}|>$ within the galaxy filament zones would reach ~ $10^{-8}$ to $10^{-9}$ G by $z = 0$. However, as Ryu, Vishniac, and others have recently calculated, large scale gravitationally-driven inflow into filaments, galaxies and galaxy groups evolve is accompanied by large scale shearing and turbulence of the IGM gas [21,22]. These models produce an amplified <|$\boldsymbol{B}_{\rm IGM}$|> in galaxy filaments that reaches as high as ~ $10^{-7}$G. Although these calculated scenarios are more removed from confirming observations, it is interesting that is can they can produce <|$\boldsymbol{B}_{\rm IGM}$|> in filaments that are comparable with the supermassive BH – generated IGM fields in equation (5). The implication for UHECR propagation and anisotropy modeling is for a high contrast in <|$\boldsymbol{B}_{\rm IGM}$|> between the filaments and voids of the local universe.

## 5. Observational evidence for $B_{\rm IGM}$ beyond galaxies and galaxy clusters.

The presence of $\boldsymbol{B}_{\rm IGM}$ can be verified by identifying faint intergalactic synchrotron radiation at the longer radio $\lambda$'s where the radiative lifetime is long enough that the synchrotron-emitting electrons can propagate far enough into the IGM from their acceleration site. This criterion favors $\lambda \approx$ 1m, and requires a surface brightness sensitivity that exceeds the capability of most current instruments.

A second method, likewise barely accessible to the best current instruments, is to detect a Faraday RM in an intergalactic context -- beyond galaxy clusters (where $\boldsymbol{B}_{\rm IGM}$ is detectable--see above). I illustrate a recent example of each method in the following 2 figures.

Figure 11 shows a 0.4 GHz continuum image of 8º diameter, minus the discrete radio sources, near the center of the Coma supercluster of galaxies [23]. The strong radio source near the center is the radio "halo" that identifies the inner Coma cluster of galaxies.

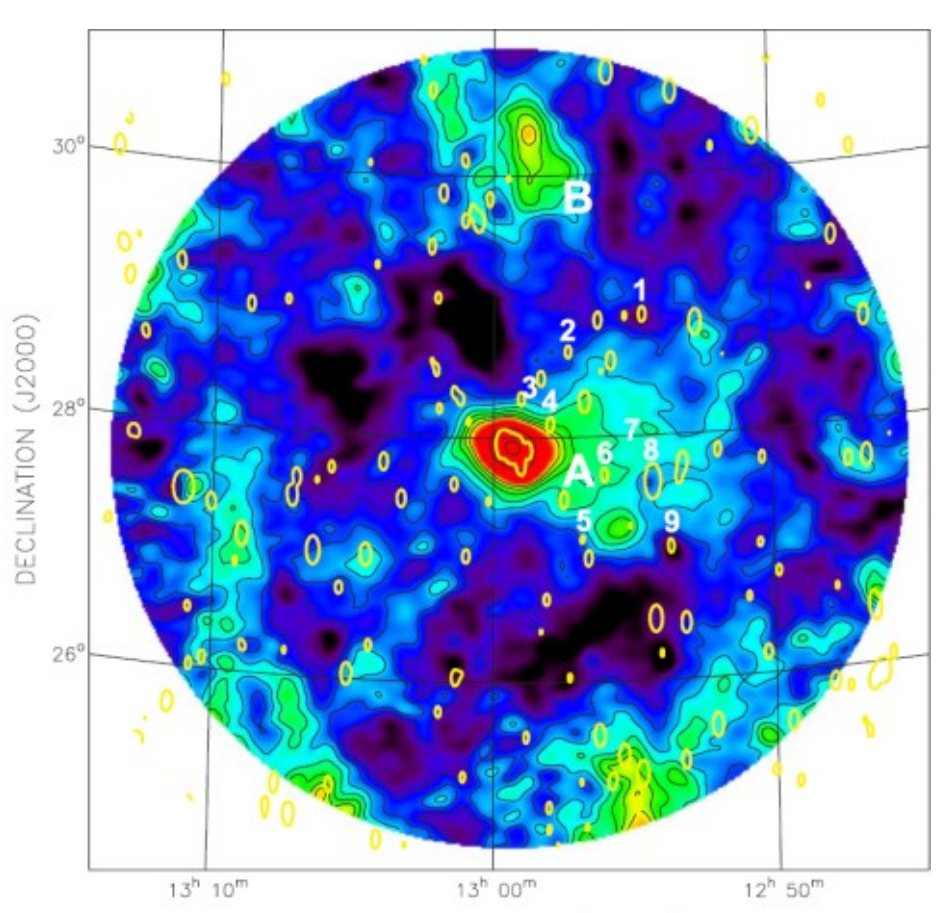

Figure 11 0.4GHz, 10' resolution image of dia. 8º showing extremely faint diffuse radio emission in the Coma supercluster of galaxies [23]. It used a combination of data from the 305m Arecibo Telescope and the NRC-DRAO Synthesis interferometer at Penticton BC. Most of the other diffuse features are not yet identified, and could be of Galactic or extragalactic origin. The rms noise level is ~250 mK at 430MHz, and is set by faint discrete source confusion.

The extended region to the right of the Coma cluster (center) is ~ 2Mpc in extent, and has intergalactic field strengths of ~ $10^{-7}$G. This is evidence for an intergalactic magnetic field strength of this order in the IGM beyond galaxy clusters.

Another recent attempt was made to detect Faraday rotation due to $\boldsymbol{B}_{\rm IGM}$ in an intergalactic filament. The well defined filament of the Perseus-Pisces supercluster zone was investigated [24] by combining appropriately smoothed Faraday rotation data with analyses of the

(a) RM- 2MASS (b) RM-CfA

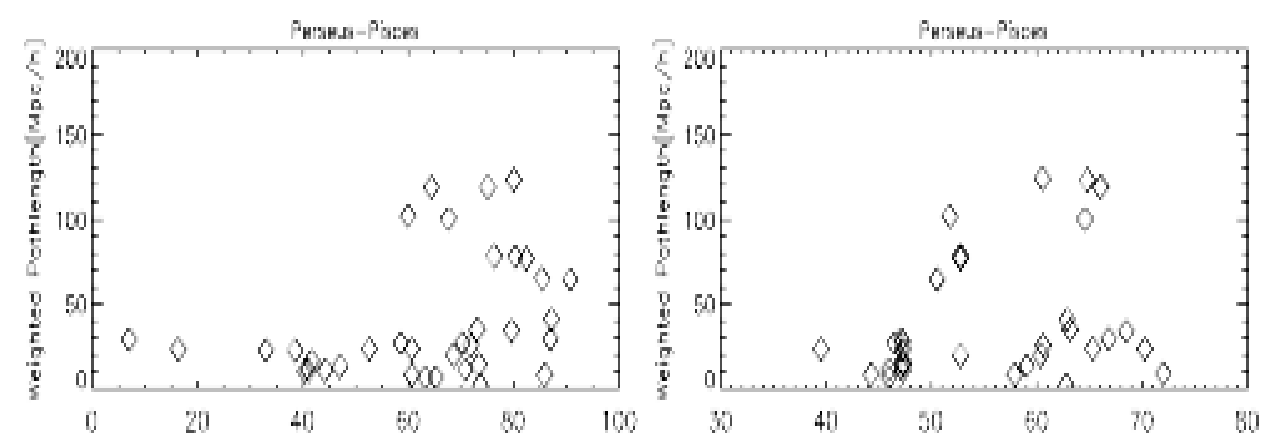

RM(rad m$^{-2}$) →

Figure 12 Plots of observed, 7º convolved RM's against calculated magneto-ionic pathlengths using two independent optical galaxy surveys, and methods described briefly in the text, and in Xu *et al*. [24]. Reproduced from [24].

2MASS and CfA surveys of galaxy magnitudes and redshifts. The smoothed 2MASS survey column densities through, the P-P supercluster filament are compared with RM in Figure 12(a). An analogous plot in Fig. 12(b) shows the RM's plotted against weighted pathlength via a 3-D Voronoi tessellation-model of the P-P filament. The results using both methods are consistent with a $\boldsymbol{B}_{\rm IGM}$ of order $10^{-8}$ – $10^{-7}$ G with a $\boldsymbol{B}$ reversal scale of a few hundred kpc. Further details can be found in [24]. This includes some model assumptions, and is always susceptible to a chance superposition of high latitude Milky Way foreground contribution to RM.

In summary, I have described a recent accumulation of evidence for intergalactic magnetic fields of $10^{-8}$ - $10^{-7}$G in galaxy filaments in the low-redshift Universe. The direct observational evidence described in this section is concordant with some magneto-plasma modeling *e.g*. [21], and direct $\boldsymbol{B}_{\rm IGM}$ from supermassive black holes *e.g*.[19]. While filament zones over Mpc pathlengths would produce strong UHECR deflections and anisotropies the fraction of such zones over a typical intergalactic propagation path appears to be small. Most of the intergalactic volume consists of voids, where $\boldsymbol{B}_{\rm IGM}$ is much smaller. Although $\boldsymbol{B}_{\rm IGM}$ in the voids is even less explored, it is very small, and can be investigated using analysis of VHECR events – a topic briefly visited below.

## 6. Exploration of voids and regions of very low $\boldsymbol{B}_{\mathrm{IGM}}$

Here it is worth recalling that the “empty’’ IGM, including the voids, is permeated by photons, by which we mean the EBL at all photon energies. Particles with energy $E_{\mathrm{P}} \gtrsim 10^{18}$ eV interact with other particles and photons, so that all of space becomes a (passive) particle physics laboratory. This fact is partially illustrated in Figure 13. At some rate, galaxy-supplied particles and magnetic fields will diffuse out of the walls and filaments into the voids. The cosmic voids are also the place to look for a pre-galactic, or primordial field. However at this point, the detection of magnetic fields in cosmic voids still lies mostly in the realm of *Gedanken* experiments.

For VHE particles the time of arrival, deflection, energy, and composition can all be measured in principle, and calculated. Figure 13, is based on ideas originally advanced by R. Plaga [25], and is a concept illustration of reactions and deflections in a intergalactic magnetic field. It is based on the existence of a burst event producing $\gamma$-rays ($\gamma$-ray burster), neutrinos, and/or hadrons, all at very high energies. The first VHE $\gamma$- rays could be products of original VHE hadrons.

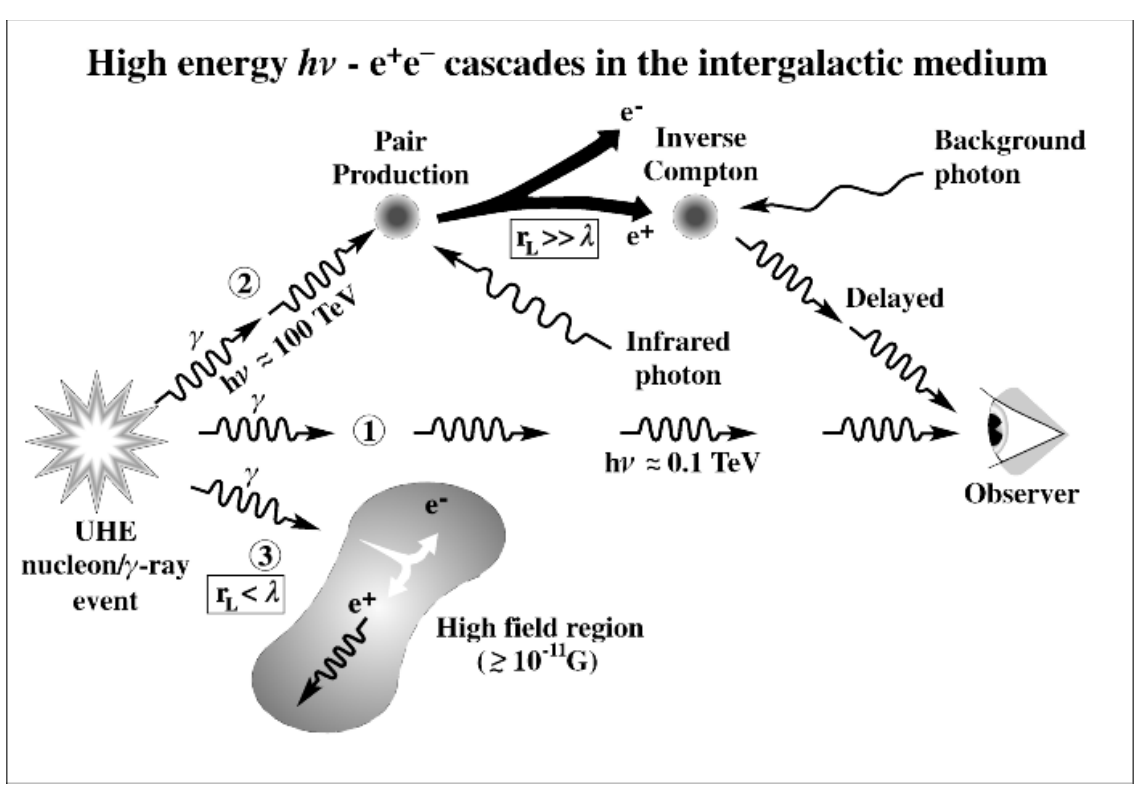

Figure 13 Propagation path 1 applies to photons of order ~ 1 Tev that propagate directly through the IGM. Photons of □ 100 TeV initiate a chain of reactions involving i.r. photons of the EBL, $e^+$, $e^-$, and, at lower energies, the CMB photons to produce a photon-particle cascade. Case 3 occurs if $\boldsymbol{B}_{\mathrm{IGM}}$ is sufficiently high to prevent the $e^+$, $e^-$ from continuing the cascade along the line of sight to us.

If the electron Larmor radii for $\boldsymbol{B}_{\mathrm{IGM}} \lesssim 10^{-12}$ G are sufficiently large, we might see a coherent photon cascade, where lower energy photons in path 1 mark the un-delayed arrival time. For $\boldsymbol{B}_{\mathrm{IGM}} \gtrsim 10^{-12}$ G we would observe a photon halo around the source of the bursting event. In this case (3 in Fig. 13), the halo size and spectrum would lead to an estimate of the magnetic field strength in the intergalactic vicinity of the source. Depending on the parameter space, the radiation could be pair annihilation radiation at 0.5 MeV, or electron/positron synchrotron radiation.

Time delays, and energy losses for intergalactic protons propagating in an intergalactic magnetic field have been calculated and discussed by Stanev *et al*.[26]. Observable effects for proton propagation here refer to a higher regime of $\boldsymbol{B}_{\mathrm{IGM}}$, in the vicinity of $10^{-9}$G. The reader could consult the interesting paper by Stanev *et al*.[26] for an informative discussion of loss mechanisms (GZK, Bethe-Heitler, etc.), and other UHECR propagation effects.

## 7 Acknowledgments

I wish to thank my many co-workers for their contributions to the material I have presented here. I thank the organizers for their invitation to participate in this very interesting meeting, and Dr Henry Glass for his invaluable assistance in the preparation of this manuscript

This work supported by the U.S. Department of Energy, to LANL, and NSERC (Canada).